# New Volleyballenes: $Y_{20}C_{60}$, $La_{20}C_{60}$, and $Lu_{20}C_{60}$

Jing Wang[a] and Ying Liu*[,a,b]

[a]Department of Physics and Hebei Advanced Thin Film Laboratory, Hebei Normal University, Shijiazhuang 050016, Hebei, China

[b]National Key Laboratory for Materials Simulation and Design, Beijing 100083, China

E-mail: <u>yliu@hebtu.edu.cn</u>

**Abstract:** New stable *Volleyballenes* $Y_{20}C_{60}$, $La_{20}C_{60}$, and $Lu_{20}C_{60}$ molecular clusters have been proposed using first-principles density functional theory studies. In conjunction with recent findings for the scandium system, these findings establish *Volleyballene* $M_{20}C_{60}$ molecules as a stable general class of fullerene family. All $M_{20}C_{60}$ ($M$=Y, La, and Lu) molecules have $T_h$ point group symmetries and relatively large HOMO-LUMO gaps.

Since the first observation of the $C_{60}$ molecule[1], known as fullerene, much effort has been invested in the study of this novel molecular cluster. In the fullerenes all the atoms are C atoms and they form a hollow sphere comprised of pentagonal and hexagonal rings. Very recently, an exceptionally stable hollow cage, composed of 20 Sc atoms and 60 C atoms, the *Volleyballene* $Sc_{20}C_{60}$ was reported. This molecular cluster has a $T_h$ point group symmetry and a volleyball-like shape.[9] The *Volleyballene* $Sc_{20}C_{60}$, as a new member of fullerene family, is receiving increasing attention[10]. If *Volleyballene* $Sc_{20}C_{60}$ does exist, it is expected that other early transition metals should be capable of forming molecules of a similar type which would also display an unusual stability. Then, we extended our work to other transition metal systems, with particular attention to elements with one *d* electron: yttrium (Y), lanthanum (La), and lutetium (Lu). Like $Sc_{20}C_{60}$, the $M_{20}C_{60}$ ($M$=Y, La, and Lu) also are found to display an enhanced stability.

The calculations were carried out with the exchange-correlation potential described by the Perdew-Burke-Ernzerhof version (PBE) of the general gradient approximation (GGA).[11] The double-numerical basis plus polarized functions (DNP)[12] was chosen. For the transition metal atoms, relativistic effects in the core were included by using the DFT semi-core pseudopotentials (DSPP).[13] All structures were fully relaxed and geometric optimizations were performed with unrestricted spin and without any symmetry constraints.

Figure 1 shows the configurations of the new *Volleyballene* $M_{20}C_{60}$ ($M$=Y, La, and Lu). The three new *Volleyballenes* $Y_{20}C_{60}$, $La_{20}C_{60}$, and $Lu_{20}C_{60}$ all have $T_h$ point group symmetries within the tolerance of 0.1 Å. Similar to the case of $Sc_{20}C_{60}$, the *Volleyballene* $M_{20}C_{60}$ ($M$=Y, La, and Lu) are composed by six $M_8C_{10}$ subunits in a crisscross pattern. In $M_8C_{10}$ subunit, two head-to-head connected carbon pentagons (C-pentagon) are surrounded by one transition-metal octagon (*M*-octagon). The 20 transition metal atoms link to form 12 suture lines. For the **$Y_{20}C_{60}$,** there are four C-C bonds with the length of 1.45 Å and one 1.46 Å in each C-pentagons. Along with a 1.485 Å C-C bond connecting the two C-pentagons, the average C-C bond length is found to be 1.455 Å. The average Y-C bond length is 2.396 Å. For the **$La_{20}C_{60}$ and $Lu_{20}C_{60}$,** the average C-C bond lengths are 1.457 and 1.452 Å, respectively. The average La-C and Lu-C bond lengths are 2.565 and 2.345 Å, respectively. For **$La_{20}C_{60}$,** both the average C-C and La-C are larger than the corresponding ones in the $Y_{20}C_{60}$, while for **$Lu_{20}C_{60}$,** they are both smaller than these in the $Y_{20}C_{60}$. All the calculated data are listed in Table I. It can be seen that the average bond lengths (C-C and

$M$-C) increased initially and then decreased as the increasing of the atomic number of transition metals.

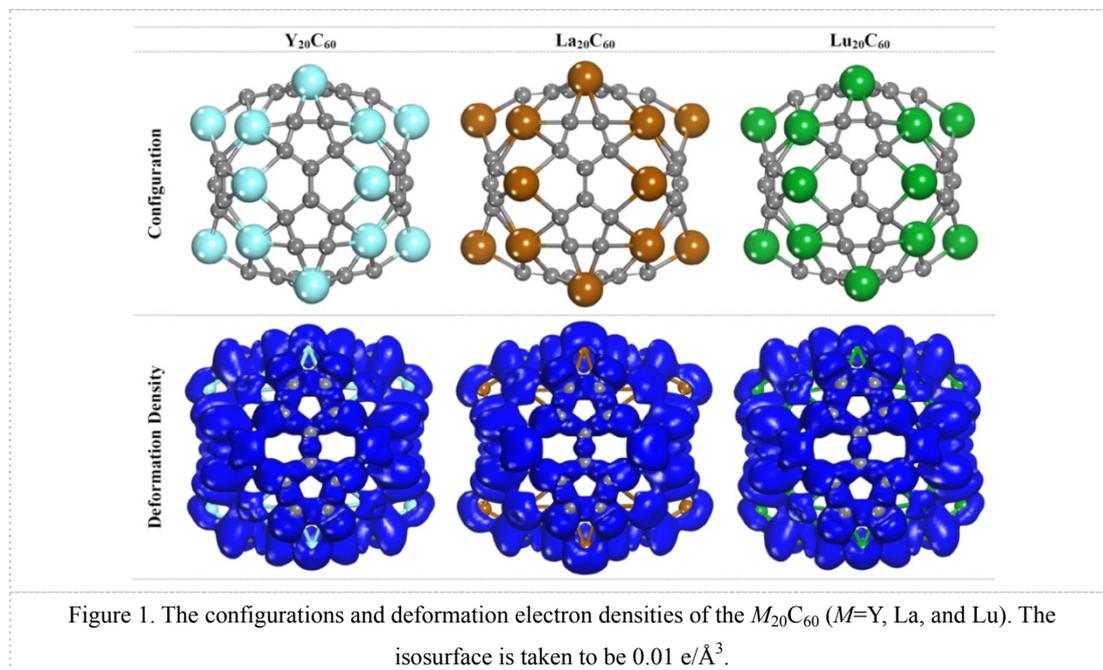

Figure 1. The configurations and deformation electron densities of the $M_{20}C_{60}$ ($M$=Y, La, and Lu). The isosurface is taken to be 0.01 e/Å$^3$.

Next, the bonding character of *Volleyballene* $M_{20}C_{60}$ ($M$=Y, La, and Lu) was investigated by analyzing their deformation electron density. The *Volleyballene* $Y_{20}C_{60}$, $La_{20}C_{60}$ and $Lu_{20}C_{60}$ have the similar bonding characteristics, mainly due to the similar electron configuration, $4d^15s^2$ for the Y atom, $5d^16s^2$ for the La atom, and $4f^{14}5d^16s^2$ for the Lu atom. In the whole, there is electron transfer from the transition metal atoms to the C atom. Mulliken population analysis shows an average charge transfer of ~0.95$e$ from each Y atom to the neighboring C atoms for $Y_{20}C_{60}$. For $La_{20}C_{60}$ and $Lu_{20}C_{60}$, the average charge transfers are ~0.69$e$ and ~0.86$e$, respectively. The transferred electrons are mainly from the $M$ $d$ state. For the C atoms, there are obvious characteristics of $sp^2$-like hybridization, and each C atom has three σ bonds. For the Sc atoms there are four lobes pointing to the neighboring four C atoms. This strengthens the link between the $M_8C_{10}$ subunits.

The stability of $M_{20}C_{60}$ was further checked using *ab* initio molecular dynamics (MD) simulations with the constant-NVE ensemble. The simulation time step was set to be 1.0 *fs* with 2000 dynamics steps. The *NVE* simulations showed that all the structures of $M_{20}C_{60}$ ($M$=Y, La, and Lu) were not disrupted over the course of a 2.0 *ps* total simulation time at an initial temperature of 1600 K, equal to a ~600 K effective temperature. Next, the vibrational frequency analysis was carried out. No imaginary frequencies were found for each of the three *Volleyballenes* $M_{20}C_{60}$ ($M$=Y, La, and Lu). These results indicate that the three new *Volleyballenes* $Y_{20}C_{60}$, $La_{20}C_{60}$, and $Lu_{20}C_{60}$ have good kinetic and thermodynamic stabilities. Figure 2 gives the calculated Raman spectra of $Y_{20}C_{60}$, $La_{20}C_{60}$, and $Lu_{20}$. The temperature here was 300 K and the 488.0 nm incident light was chosen in order to simulate a realistic Raman spectrum that can be compared to experimental results.

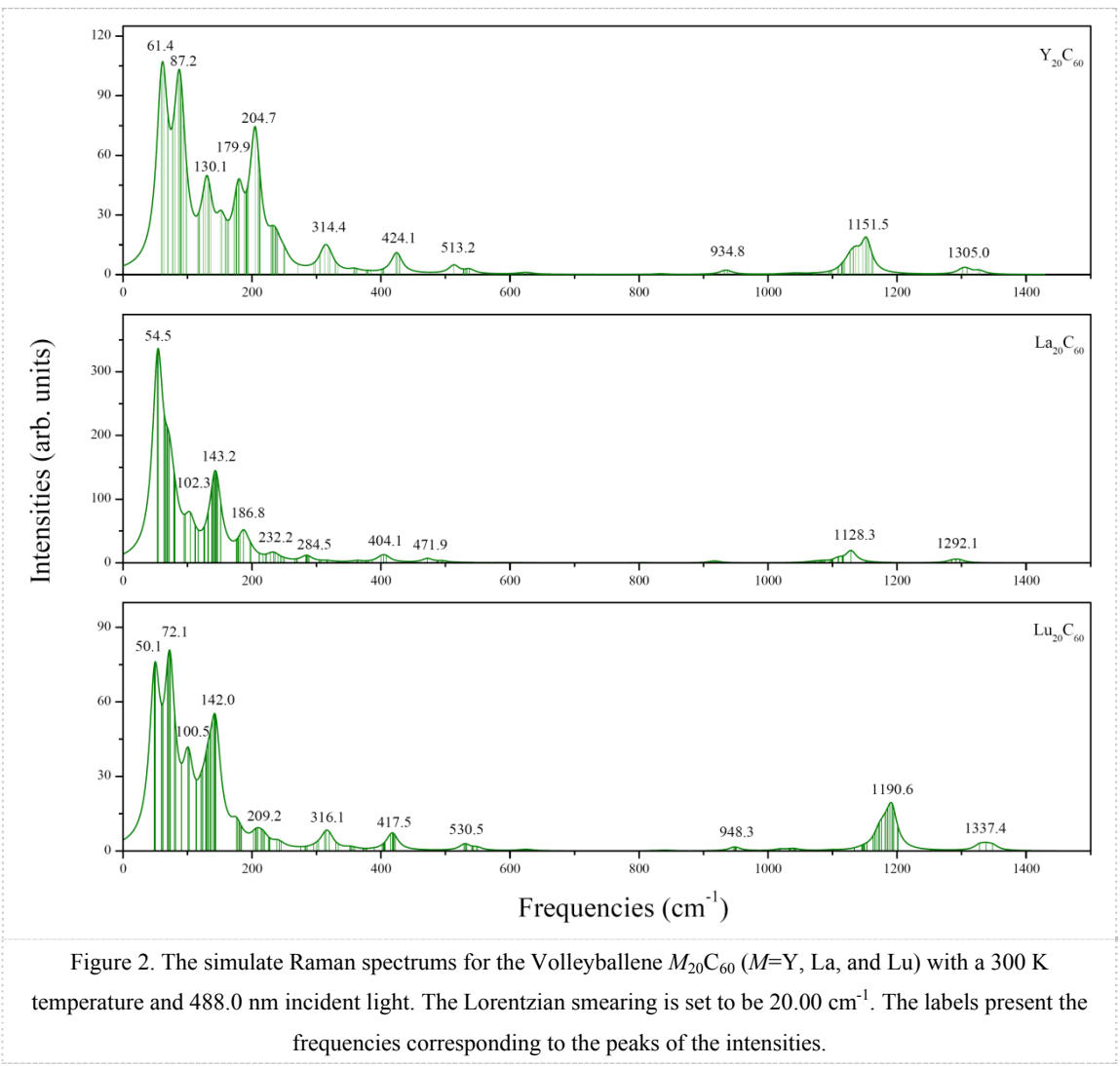

Figure 2. The simulate Raman spectrums for the Volleyballene $M_{20}C_{60}$ (*M*=Y, La, and Lu) with a 300 K temperature and 488.0 nm incident light. The Lorentzian smearing is set to be 20.00 cm$^{-1}$. The labels present the frequencies corresponding to the peaks of the intensities.

Then, we calculated the partial density of states (PDOS) and frontier molecular orbitals, including the highest occupied molecular orbital (HOMO) and the lowest unoccupied molecular orbital (LUMO) of $Y_{20}C_{60}$, $La_{20}C_{60}$, and $Lu_{20}C_{60}$ as shown in Figure 3. From the contours of the HOMO orbitals, it can be seen that the HOMO orbitals are mostly localized on the C atoms. There is also obvious s*p-d* hybridization. As for the case of LUMO, the energy level features show that the LUMO orbital is doubly degenerate. On part of the transition metal atoms, there are obvious $d_{z^2}$-like orbitals characteristics. Close examination of the partial density of states (PDOS) further confirms the hybridization characteristics of the HOMO and LUMO orbitals. All these results demonstrate that the hybridization between *M d* orbitals and C *s-p* orbitals is essential for stabilizing the cage structure of $M_{20}C_{60}$ (*M*=Y, La, and Lu).

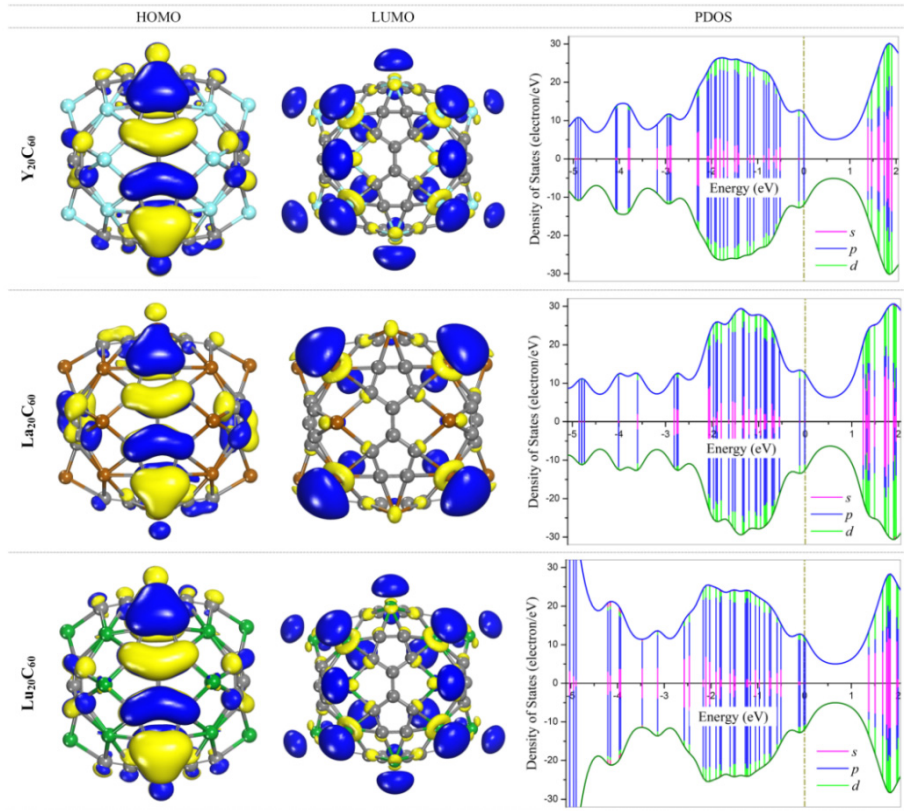

Figure 3. The HOMO and LUMO orbitals, and PDOS for the $M_{20}C_{60}$ (M=Y, La, and Lu). The isosurface for the orbitals is set at the value of 0.015 e/Å$^3$.

For the *Volleyballenes* $Y_{20}C_{60}$, $La_{20}C_{60}$, and $Lu_{20}C_{60}$, relatively large HOMO-LUMO gaps were observed as listed in Table I. Among them, $Y_{20}C_{60}$ has the largest HOMO-LUMO gap and with the increase of the atomic number, the HOMO-LUMO gaps first decreases and then increases. The large gaps are mainly due to the energy of the *d* atomic orbitals being much lower than that of the *p* orbitals. Therefore, the three new *Volleyballenes* $Y_{20}C_{60}$, $La_{20}C_{60}$, and $Lu_{20}C_{60}$ should be stable buckyballene variants with exceedingly high chemical stability.

Table I. Summary of the calculated results for $M_{20}C_{60}$ (*M*=Y, La, and Lu). The data include the symmetry group (Sym.), the average C-C ($d_1$) and *M*-C ($d_2$) bond lengths, the average charge transfer from *M* to carbon atoms, and the HOMO-LUMO energy gap ($E_g$) in the units of Å for the length and eV for energy.

|  | Sym. | $d_1$ | $d_2$ | Q | $E_g$ |
|---|---|---|---|---|---|
| $Y_{20}C_{60}$ | $T_h$ | 1.455 | 2.396 | 0.953 | 1.395 |
| $La_{20}C_{60}$ | $T_h$ | 1.457 | 2.565 | 0.693 | 1.254 |
| $Lu_{20}C_{60}$ | $T_h$ | 1.452 | 2.345 | 0.856 | 1.378 |

The first-principles studies have identified a new series of stable *Volleyballenes*, $Y_{20}C_{60}$, $La_{20}C_{60}$, and $Lu_{20}C_{60}$. In the initial report on the stability of $Sc_{20}C_{60}$[9], we speculated that $Sc_{20}C_{60}$ may comprise one member of a *Volleyballene* family, and that other transition or rare-earth metals could also form stable $M_{20}C_{60}$ molecular clusters.

*Acknowledgements*

This work was supported by the National Natural Science Foundation of China (Grant Nos. 11274089, 11304076, and U1331116), the Natural Science Foundation of Hebei Province (Grant No. A2012205066), and the Science Foundation of Hebei Education Department for Distinguished Young Scholars (Grant No. YQ2013008). We also acknowledge partial financial support from the 973 Project in China under Grant No. 2011CB606401.